\documentclass[aps,showpacs,nofootinbib]{revtex4}

\usepackage{graphicx}
\usepackage{graphics}
\usepackage{amssymb}
\usepackage{amsmath}
\usepackage{bm}

\newcommand{\be}{\begin{equation}}
\newcommand{\ee}{\end{equation}}
\newcommand{\bea}{\begin{eqnarray}}
\newcommand{\eea}{\end{eqnarray}}
\newcommand{\beal}{\begin{align}}
\newcommand{\eal}{\end{align}}
\newcommand{\bespl}{\begin{split}}
\newcommand{\espl}{\end{split}}

\newcommand{\nslash}{\kern 0.2 em n\kern -0.50em /}
\newcommand{\kslash}{\kern 0.2 em k\kern -0.45em /}
\newcommand{\pslash}{\kern 0.2 em p\kern -0.50em /}
\newcommand{\Sslash}{\kern 0.2 em S\kern -0.50em /}
\newcommand{\Pslash}{\kern 0.2 em P\kern -0.50em /}
\newcommand{\Rslash}{\kern 0.2 em R\kern -0.50em /}

\begin{document}

\title{T-odd distribution functions, breaking of long range 
correlations, and sudden entropy changes, 
in Drell-Yan high-energy processes.
}

\author{A.~Bianconi}
\email{andrea.bianconi@bs.infn.it}
\affiliation{Dipartimento di Chimica e Fisica per l'Ingegneria e per i 
Materiali, Universit\`a di Brescia, I-25123 Brescia, Italy, and\\
Istituto Nazionale di Fisica Nucleare, Sezione di Pavia, I-27100 Pavia, Italy}

\begin{abstract}
Time-Odd parton distribution functions in a Drell-Yan 
process are here 
studied by examining the evolution of the 
internal statistical properties of the 
interacting hadrons. Time-Odd functions are 
shown to be a signature of the irreversible 
process in which a hadronic state characterized by long range correlation 
properties (hadronic phase) 
decays to produce a cloud of independent partons (partonic phase) 
because of initial/final state interactions. 
The relevant considered variable is the rate of increase 
of the entropy of the hadronic system. This quantity is 
shown to be roughly equal to the 
decay rate of the 
hadronic state. Conditions for getting a leading twist  
Time-Odd effect are 
established on this basis. 
Last, the relevant case of a large entropy increase  
associated with transverse-dominated 
initial/final state interactions 
is analyzed. 
\end{abstract}


\maketitle

\section{Introduction}

The field of Time-reversal odd (T-odd) distribution 
functions in high energy hadronic physics has known a rich development  
in the last years, and these distributions have become 
an essential tool in the intepretation of the phenomenology 
of transverse spin and azimutal asymmetries in high energy physics, 
and in the design of new experiments. 

In the framework of QCD factorization the existence was predicted 
long ago\cite{EfremovTeryaev82,QiuSterman91} of twist-3 T-odd effects. 
The first T-odd leading twist distribution function was 
the Sivers function\cite{Sivers} in 1990, 
followed by the 
Boer-Mulders-Tangerman 
function\cite{MuldersTangerman96,BoerMulders98,Boer99}. 
The Sivers function was used to explain single spin 
asymmetries\cite{ABM95}, the BMT function to explain 
unpolarized Drell-Yan azimuthal 
asymmetries\cite{Boer99}. These phenomena could also be interpreted 
via soft mechanisms\cite{BorosLiangMeng}, despite still 
according with schemes that 
may be (qualitatively) considered ``T-odd''. 
It has been recently shown\cite{JQVY06} that 
in processes like Drell-Yan where the scale of the 
transverse momentum $q_T$ is independent from the hard scale $Q$, 
the twist-3 T-odd structures predicted 
in \cite{EfremovTeryaev82,QiuSterman91}
are not $Q-$suppressed at finite $q_T$, and they 
essentially coincide with the above quoted leading twist T-odd 
distributions. 
That work also shows that the details of the initial state may 
be of secondary relevance when the problem of producing leading 
twist T-odd distributions is addressed. 

A time reversal 
odd structure function, the so-called ``fifth structure function'',  
was also long ago introduced\cite{Donnelly} 
and more recently modeled\cite{BB95,BR97} in nuclear physics to 
describe normal asymmetries 
in $A(e,e'p)(A-1)$ quasi-elastic scattering 
(for reviews see \cite{BGPR}). 
It was known\cite{Donnelly} that this structure function could 
exist in presence of final state interactions only, at the point that 
its vanishing was suggested\cite{BB95} as an experimental signature 
for the onset of Color Transparencey.
The generalization of nuclear physics experience 
to high energy hadronic physics cannot be complete 
for the absence of a factorization framework 
in the former case. One interesting and perhaps more general 
point, however, is that 
despite the phenomenology of the fifth structure function is 
strictly related with spin asymmetries, models for it\cite{BB95,BR97} 
do not need to involve spin-dependent interactions. Rather, 
the key point is in particle flux absorption. 

In QCD, leading twist T-odd distribution functions 
were initially considered forbidden\cite{Collins93} by general 
invariance principles. 
After an explicit mechanism was shown to produce a 
nonzero T-odd 
distribution in a QCD framework\cite{BrodskyHwangSchmidt02}, 
the existence problem was systematically fixed in 
ref.\cite{Collins02}, where Collins observed that the presence 
of a link operator necessary to restore gauge invariance for 
the two-point correlation function allowed for non-vanishing T-odd 
distributions in QCD. These and other 
works\cite{JiYuan02,BelitskyJiYuan03} 
also give 
conditions for having leading twist effects in 
$k_T-$dependent T-odd functions.  

Models\cite{Yuan03,GambergGoldsteinOganessyan03,BoerBrodskyHwang03,
BacchettaSchaeferYang04,LuMa04} 
or studies\cite{Pobylitsa03,DalesioMurgia04,Burkardt04,
Drago05,GoekeMeissnerMetzSchlegel06,BoerVogelsang06} 
around T-odd distribution functions have been 
produced and discussed by several 
authors. Recently several phenomenological 
parameterizations 
of the Sivers 
function for quarks\cite{Torino05,VogelsangYuan05,BR06a,
CollinsGoeke05} and gluons\cite{AnselminoDalesioMelisMurgia06} 
have been deduced 
from available 
data\cite{Star,Hermes,Compass,Phenix}. 
An old parameterization\cite{Conway89} 
of the $cos(2\phi)$ asymmetry has been translated into 
parameterizations\cite{Boer99,BaroneLuMa05} 
of the Boer-Mulders-Tangerman function. 
Several experiments aimed at the measurement of T-odd functions 
are planned for the next ten 
years\cite{panda,assia,pax,rhic2,compassDY}. 

In this work the problem of the existence of leading twist T-odd 
functions and of their interpretation is examined from an 
unusual point of view. 
They are supposed to 
originate in a phase transition of the 
statistical properties 
of the hadron ground state due to initial/final 
state interactions accompanying a hard electromagnetic process. 

With ``phase'' we only mean a set of collective 
properties enjoying a relative stability. We may name 
``hadronic'' phase the initial one, and ``partonic'' phase 
the final one. For having leading twist 
observables, the transition 
needs to take place over a time range that is singular in 
the infinite momentum limit. 

We will here focus on the Drell-Yan process only, so ``partonic 
distributions'' means actually ``partonic distributions measured in 
Drell-Yan''. 
On the ground of eclipse effects, one would expect 
Drell-Yan 
rates on nuclear targets to scale as $A^{2 / 3}$, possibly with 
shadowing effects on the shape of the measured distributions. 
The situation ir rather different 
(see \cite{Kenyon82} for a detailed 
compared review of Drell-Yan proton and pion data on light to heavy 
targets). Data 
show that hadron/nuclear matter is 
practically transparent to individual partons, while hadronic 
states 
have a rather short survival path in it. The 
partonic distributions resulting from a Drell-Yan measurement 
are independent from the target mass number $A$  
from H to Pt, and the total Drell-Yan 
rate is $\propto$ $A$, while the cross sections 
for the much more frequent hadronic processes resulting from 
the same collisions are $\sim$ $A^{2/3}$ as 
obvious. Drell-Yan data recover the $A^{2/3}$ scaling law 
if the dilepton pair is on some resonant mass like $\rho$ 
or $J/\psi$. 

This suggests that any single partons is able to 
transport $probabilistic$ information on its initial 
state through all the volume of a heavy nucleus. On the contrary, 
its initial state wavefunction is well known to be destroyed 
within 1 fm. 
To say that hadron-hadron initial state 
interactions cause a transition from a hadronic 
to a partonic phase, destroying the long range properties 
of the former, but without touching the short range ones 
associated with the latter, is just another way to say 
the same thing with other words. \footnote{ 
The 
words ``partonic phase'' do not refer specifically to 
quark-gluon states like 
those studied in the physics of heavy ions 
collisions (see e.g. \cite{CernYR2004} and references therein). 
They refer more generically to any hadron state 
where long range correlations are not present. 
}

However, this way of seeing 
the process will be useful in the following. In particular, it 
suggests a study of the entropy properties of the system, since 
any process where long range correlations are atomized 
is associated with a relevant increase of entropy. 
T-odd functions are here considered as a direct signature 
of the hadron to parton transition, and a quantitative 
measure of the time rate of increase of the entropy. 

A weak point of the picture presented here can obviously be 
the difficulty in disentangling specific effects of QCD/strong 
interactions from general properties of a bound system subject 
to a hard external probe. 

Section II is devoted to listing a few general  
definitions and relations that are systematically 
used in the following. 

The presence of chaotic processes in association with 
hard hadronic interactions is 
obvious at a generic level. 
More specifically however we need 
to understand where the chaotic side of these processes 
enters formally into parton distribution functions, 
how this leads to T-odd observables, and at which conditions 
these observables are leading twist. 
To this point sections III and IV are devoted. In section III 
the correlator is factorized into two parts. One contains 
those effects like S-P interference that are specific of the 
detection process, while the other one contains the statistical 
properties (``2-scalar correlator''). It is 
shown that the 2-scalar correlator may be 
decomposed with completeness in damped plane waves of the form 
$exp(-ix\xi - \gamma(x) \xi)$, where $\xi$ $\equiv$ $P_+ z_-$,  
and the condition of finiteness 
of $\gamma(x)$ is shown to be a signature of leading twist 
T-odd effects. 
In section IV a review of some topics of statistical 
mechanics is given, focussing in particular on 
the decay properties of a self-correlation function, and 
on the correspondence with the relevant quantities  
of the Drell-Yan problem. The general criterion 
$\gamma$ $\sim$ 
$dS/d\xi$ $\sim$ $x$ relating the rate of change of 
entropy with the hadron state decay rate $\gamma$ 
and with 
the existence of T-odd leading twist effect 
is qualitatively introduced.

In section V the previous arguments are made more precise 
and specific. First, a criterion is given to somehow 
formalize the breaking of a long range correlation, i.e. 
to establish a borderline between long and short range 
correlations. Then, the relevant phase space entering 
the definition of entropy for long range and short range 
states is focussed. Initial state 
interactions causing the breaking 
are divided into two classes 
(single and multiple event processes) that are separately 
examined. 

Section VI is devoted to a more restricted group of processes.  
Let $A(+)$ and $A(-)$ be the 
``blobs'' graphically associated with factorization separated 
areas of the full process. 
A peculiar side of linking T-odd effects with entropy changes 
is that it is natural to build models where an arbitrarily 
large increase of the entropy of $A(+)$ takes place in absence of 
large energy exchanges between $A(+)$ and $A(-)$. Hard exchanges 
of transverse momentum between $A(+)$ and $A(-)$ may trigger large 
energy transferss among degrees of freedom 
all belonging to $A(+)$ alone. This is relevant since 
hadron-hadron interactions associated with large energy exchanges 
between the hadrons 
are statistically suppressed at large energies. So it is fair to 
imagine effective initial state interactions to be dominated by 
exchange of trasverse momentum. In addition, 
energy conservation allows for a more precise calculation 
of the phase space and of the entropy.

\section{Some general definitions and notations} 

We consider a Drell-Yan process taking place between two colliding 
hadrons with momenta $P$ and $P'$. The parton we will examine in the 
following is a quark belonging to the former hadron. 

To select leading twist terms, we will use the infinite 
momentum limit $P_+$ $\rightarrow$ $\infty$, for $P_+$ 
measured in the 
center of mass frame of the two colliding hadrons. 
In other words, $(P + P')^2$ $\sim$ $P_+P_-'$ 
is arbitrarily large in this limit. 

We name $A(+)$ and 
$A(-)$ those 
areas of the process/diagram associated with 
factorization-separated kinematical processes. 
The area of our interest is 
$A(+)$, where the relevant components are 
and $k_+$ (momentum), and 
$z_-$ (spacetime). 
All those fields whose momentum 
component $p_+$ 
is $O(P_+)$ belong to $A(+)$. 
We will ignore completely what happens 
inside $A(-)$. 
We will however consider 
4-momentum exchanges between $A(+)$ and $A(-)$, 
potentially undermining factorization. 

The relevant 
process considered in this work is the one 
corresponding to the 
traditional cut amplitude associated by 
second order unitarity to the inclusive hadronic 
distribution functions: 

In this process a hole is created in a hadron state 
by extracting a quark/antiquark in a point $z_\mu'$, the 
``hadron+hole'' set propagates through real states only, 
and in $z_\mu''$ the hole is filled and the original hadron 
state is restored 
by reinserting the quark in its place. During the propagation, 
the hole is subject to a gauge restoring field according to 
ref.\cite{Collins02}. 
We will sometimes speak of 
``quark interactions'' actually meaning ``quark hole 
interactions''.  
\footnote{E.g., in spectator models the hole path starts from 
the quark-photon vertex, 
runs backwards along the quark propagator up to the 
hadron-quark-diquark vertex, follows the diquark across the 
cut up to the other hadron-quark-diquark vertex, 
and next a quark propagator up to the other photon-quark 
vertex.} 

The extracted quark has momentum 
$k_\mu$ with lightcone 
$k_+$ $\equiv$ $x P_+$ and negligible $k_-$. The impact 
parameter $\vec z_T$ is the space vector conjugate to $\vec k_T$. 
The spacetime displacement of the hole from $z_\mu'$ to $z_\mu''$ 
is $z_\mu$ $\equiv$ $z_\mu'-z_\mu''$ 
$\equiv$ $(z_-,0,\vec z_T)$ $\approx$ $(z_-,0,\vec z_T)$. 
The fourth coordinate $z_+$ plays no role and 
is not explicitely reported in the following. 

Any relevant distribution function $q(x)$ is the 
Fourier transform 
with respect to $z_-$ and $z_T$ of a correlation function 
$g(z_-,\vec z_T)$. 
Leading twist effects are naturally selected by writing the 
Fourier transform with respect to the scaled variables 
$x$ and $\xi$: 

\begin{equation}
\xi\ \equiv\ P_+ z_-, \label{eq:def_xi}
\end{equation}

\begin{equation}
q(x,\vec k_T)\ \equiv\ P_+ \int e^{-i xP_+z_-} e^{ik_T z_T}
g(z_-,\vec k_T) dz_- d \vec k_T\ \equiv\ 
\int e^{-i x\xi} e^{ik_T z_T}
G(\xi,z_T) d\xi. \label{eq:def_qx}
\end{equation}

$G(\xi,z_T)$, $\xi$, $x$, etc will be named ``scaled'' quantities, 
$g(z_-,z_T)$, $z_-$, $k_+$, etc 
``non-scaled'' ones. 
The range of useful values of $\xi$ remains 
finite $\sim$ $1/x$ when $P_+$ $\rightarrow$ $\infty$. 
So in this limit 
the function $G(\xi)$ does 
not need to be singular in the origin to produce a nonzero $q(x)$.  
So, in most of the following the discussion is in terms of the 
scaled quantities. 
On the other side, it will be useful sometimes to get back 
to the non-scaled $z_-$ representation to evidentiate singularities. 

In all cases, $g$ or $G$ are here assumed as complex numbers and not matrix 
objects, so they must be read as a 
projection of the spinor correlation matrix $g_{ij}$ over a suitable 
operator, e.g. $\gamma^+$. The $\gamma^+$ projection is the 
most confortable one, since it selects simultaneously the main 
unpolarized 
quark distribution and the Sivers T-odd function. To get the former 
one needs to sum over opposite transverse polarization states, 
to get the latter one takes the difference, so e.g. in this case 
$G$ means $G_+ \pm G_-$. More in general, $G$ must be read as 
a linear combination of correlator projections, suitable to project 
interference terms between even and odd angular momemtum waves. 

For simplicity of the discussion 
we will assume that the transverse angular dependence is 
factorized in the correlator: 
\begin{equation} 
G(\xi,\vec z_T)\ \equiv\ G(\xi,z_T)G_A(\hat z_T). 
\label{eq:def_g_a}
\end{equation} 

\begin{equation} 
q(x,\vec k_T)\ \equiv\ q(x,k_T)q_A(\hat k_T). 
\label{eq:def_q_a}
\end{equation} 


In general, for a given hadron spin along $\hat x$ 
we have a double partial 
wave sum 
$G(z_\mu,0)$ $=$ 
$\sum_{L,M,L'}g_{L,L'}Y_{L,M}(\theta_x,\phi_x)Y_{L0}(0)$.  
Eqs.\ref{eq:def_g_a}
and \ref{eq:def_q_a} refer to one term only in the sum. 


In eq.\ref{eq:def_g_a} 
the former term will be named ``2-scalar'' 
correlator, the latter 
``angular'' correlator. The word ``2-scalar'' only refers to 
2-dimensional transverse rotations, not to general spacetime 
properties. Distribution functions are 
defined factorizing $\hat k_T$-dpendence out of them,  
so they are in $q(x,k_T)$. Despite in general 
we have a double sum 
of partial wave contributions,  
for the following discussion one term (clearly an interference 
term between different angular momenta) 
is quite enough. 

The role of the angular parts is discussed in next section. 
Apart for that section, 
in the rest of the paper 
the discussion regards the 2-scalar correlator and distributions only, 
and the word ``2-scalar'' is omitted. 
Since in the 2-scalar correlator 
the explicit presence of $z_T$ and $k_T$ is often not 
necessary, we just write $G(\xi)$ and $q(x)$ in all the relations 
where $k_T$ is not explicitly needed. 

\section{Imaginary part of the 2-scalar correlator} 

Having defined $q(x,\vec k_T)$ as the Fourier transform of an 
amplitude 
that is projected on intermediate 
$real$ states only, $q(x,\vec k_T)$ is the imaginary part of 
a complex amplitude,  
and as such it is real. 
Since however only $q(x,\vec k_T)$ is bound to 
be real, we may have 

\begin{equation} 
q(x,\vec k_T)\ =\ Re[q(x,k_T)] Re[q_A(\hat k_T)]\ -\ 
Im[q(x,k_T)] Im[q_A(\hat k_T)]. 
\label{eq:imaginary_product}
\end{equation} 

Clearly, if $q(x,k_T)$ represents a single partial wave contribution 
to the full distribution, only one of the two terms in the 
right-hand side will be nonzero. 
For an angular even-odd interference 
term, the nonzero one should be the Imaginary-Imaginary one. 
Since this statement is not completely free from ambiguity, I take 
it as a simplifying assumption. 
Looking for T-odd 
contributions, we will focus on this term. 

A leading twist T-odd effect manifests itself in 
a finite $Im[q(x,k_T)]$. Let us make this point more precise, 
following the scheme adopted in ref.\cite{BR97},  
and suppose that 
some of the Fourier components of a T-even $G(\xi)$ 
are substituted by damped waves 

\begin{equation}
e^{ix\xi}\ \rightarrow\ e^{i x\xi - \gamma \xi}
\label{eq:damped_wave}
\end{equation} 
where $\gamma$ in general depends on $x$. 
In other words, the integration path is shifted from the real axis 
and the Fourier transform becomes a Laplace transform. 
For the Laplace transform general theorems exist, but here we would 
like to rely on physics first, and demonstrate that the above waves 
are a proper and complete set of eigenfunctions for a decomposition 
of $G(\xi)$. 

The above damped waves correspond to a nonconserved 
decaying probability at increasing times, and are 
eigenfunctions of a hamiltonian operator 
that enjoys two relevant properties: 

(i) It is non-hermitian and so it does not respect time 
inversion symmetry. 

(ii) It is invariant for $\xi-$translations. 

In addition, we must notice that the above damped waves are the 
only possible eigenfunctions for a hamiltonian that 
satisfies both (i) and (ii). 

An amplitude $G(\xi',\xi'')$ describing the hole propagation 
under the action 
of the gauge restoring field is not 
invariant for $\xi-$translation, so it depends  
on both $\xi'$ and $\xi''$ and not on their difference $\xi$ 
only. Its Fourier transform depends on 
$two$ $x$ values corresponding to the incoming and outcoming 
momenta. 

Since by definition only $x-$diagonal 
states may be present in a distribution function, 
these states 
must be eigenstates of a $\xi-$homogeneous and non-hermitian 
Hamiltonian. 
Non-hermiticity depends on the fact that it includes 
the gauge restoring field, but at the same time it 
excludes those eigenstates 
that, because of this field, do exist and contain 
violation of $x-$conservation. 

Let us define $\hat H_{true}$ as the true full hamiltonian 
including the gauge restoring field. $\{F\}$ is the full 
set of eigenstates of $\hat H_{true}$, and 
$F$ $\equiv$ $\{XC\}$ $+$ $\{XNC\}$, 
where $\{XC\}$ is the set of $x-$conserving states and $\{XNC\}$ its 
complementary. The hamiltonian $\hat H$ that one is really 
using is the projection of $H_{true}$ on $\{XC\}$ 
and such a projection is well known to 
have complex energy eigenvalues (all the theory of the 
nuclear optical potential is based on this fact and presents 
close analogies with what is discussed here, see 
e.g.\cite{Feshbach}). On the other side, the projection 
is by definition $x-$diagonal and so invariant for 
$\xi$-translations. 

Summarizing, 
$G(\xi)$ describes the hole propagation according to a 
Hamiltonian that is (i) non-hermitian, (ii) $\xi-$homogeneous. 
For these reasons  
it must admit a complete 
decomposition in states of the above form eq.\ref{eq:damped_wave} 

\begin{equation}
G(\xi)\ \equiv\ \int_0^1 dx' e^{i x \xi} G(x')
\ \rightarrow\ 
\int_0^1 dx' e^{i x' \xi - \gamma \xi} G(x'). 
\label{eq:laplace}
\end{equation} 

The Fourier transform of each damped wave with respect to 
$exp(-ix\xi)$ introduces the factor 

\begin{equation}
{1 \over {x - i\gamma}} 
\label{eq:damped_pole}
\end{equation} 
with imaginary part $2\gamma/(\gamma^2 + x^2)$
into the Fourier 
transform of the 2-scalar 
correlator. The ratio of the imaginary to the real part is 
$\gamma/x$. 

The damping $\gamma$ must be finite: 
an infinitesimal damping of plane waves is normally assumed to 
avoid convergency problems without introducing T-odd effects. 
In the infinite momentum limit $k_+$ $\equiv$ $x P_+$, 
$P_+$ $\rightarrow$ $\infty$, this requirement is rather strong, 
since it implies that in the non-scaled spacetime we have a 
damped wave $exp(i xP_+ z_- - \Gamma z_-)$ with infinite $\Gamma$. 
This is however necessary, since the Fourier transform 
$exp(-i xP_+ z_-)$ effectively 
probes regions of size $\Delta z_-$ $\sim$ $1/xP_+$, and within 
this $z_-$ range the damping is negligible unless 
$\Gamma$ is $O(xP_+)$. 

So, the condition for a leading twist effect, largely used in 
the following, is 
\begin{equation}
\gamma(x)\ \sim\ x,\ for\ P_+\ \rightarrow\ \infty
\label{eq:gamma1} 
\end{equation} 
equivalent to $\Gamma$ $=$ $O(xP_+)$. 

On the contrary, a finite $\Gamma$ ($\rightarrow$ $\gamma$ 
$\sim$ $1/P_+$) allows for a finite T-odd effect 
as far as $P_+$ is finite. 
In other words, one gets a higher twist effect.

As above observed, 
$q(x,\vec k_T)$ must be real, so in eq.\ref{eq:imaginary_product} 
$Im[q(x,k_T)]$ must combine with a nonzero $Im[q_A(\hat k_T)]$. 
In all the models known to the author, T-odd properties become  
observable in 
interference terms between odd and even waves of 
orbital angular momentum. Typically, 
between S and P waves. 

To avoid ambiguities, we remark that 
in the scheme adopted in this work, $two$ kinds 
of interference effects have relevance: 

(i) interference between 
continuous sets of states (due to initial state interactions) 
producing an imaginary part in the Fourier transform of the 
2-scalar correlator, 

(ii) interference 
between a few well identified angular waves, producing an imaginary 
part in the Fourier transform of the angular correlator. 

Missing one of the two, T-odd effects 
are not observable. Without interference terms 
the angular correlator is even with respect to $\vec z_T$, 
so its Fourier transform is real. 
A well chosen interference between even and odd angular waves, 
quantized with respect to the $\hat y$ axis, 
may imply 
the simultaneous presence of a $z_x$-even and of $z_x$-odd term 
in $G_A(\vec z_T)$. 
This produces both a real and an 
imaginary part in the $\vec k_T$-Fourier transform of $G_A(\vec z_T)$.

\section{Non-adiabatic decay of a self-correlation function} 

\subsection{Irreversibility} 

In the physics of the strong and electromagnetic interactions, 
no phenomena exist that may cause, at fundamental level, a time 
asymmetry. This situation 
is conversely 
quite common when from the fundamental level one passes to 
(i) systems where a huge number degrees of freedom are involved, 
(ii) systems where some degree of freedom is hidden in a 
non-hermitian hamiltonian, (iii) degrees of freedom whose 
time frequencey spectrum is continuous. The first class  
is a special case of the last one since strict irreversibility 
is obtained when the number of degrees of 
freedom tends to infinity. 
The second class also is a special case of 
the last one, since a non-hermitian hamiltonian has complex 
energy eigenvalues, implying a finite width for energy levels. 
So we will adopt the last and most general definition  
(seemingly introduced by N.S.Krilov, see\cite{Krilov} for 
a later recollection of his works in English) 
of an irreversible process. 

\subsection{Fok-Krilov theorem}

In nonrelativistic quantum mechanics 
the decay of a self-correlation function can be present 
when a system $A$ described by a stationary 
wavefunction $\psi_o(x,t)$ 
starts interacting with another system $B$, its wavefunction 
$\psi(x,t)$ deviates from the $\psi_o$ form and 
looses its previous 
stationarity properties. It can be generically written 
as a sum over eigenstates of the full hamiltonian (including 
interactions) with eigenvalues $E$. 
If the sum is over a continuous $E$ set, 
the self-correlation function 
\begin{equation}
C(t)\ =\ \int dx \psi^*(x,t)\psi_o(x,0) 
\label{eq:selfcorrelation1}
\end{equation}
is a decaying function of $t$ for $t$ $\rightarrow$ $\infty$. 
Else, the correlator is periodic 
(Fok-Krilov theorem\cite{FokKrilov47},  
see e.g. ref.\cite{Davidov} for a 
simple introduction). 

The meaning of the self-correlation function is just the 
overlap of a state with itself at a later time. 
If the correlator decays, it means that 
the evolved wavefunction has no resemblance with its initial 
form. 
Clearly, T-odd processes are associated with decaying correlators 
and irreversible processes. 

\subsection{Entropy}

When a previously discrete energy spectrum $\{E_{on}\}$
is turned continuous,  
we may define a new fictitious set $\{E_n\}$ of energy levels 
such that any $E_n$ is a function of $E_{on}$ and the 
corrispondence is one-to-one. 
Then we may write the true continuous energy eigenvalues $E$ 
in the form 
$E$ $\equiv$ $E_n + a(E_{n+1}-E_n)$, 0 $\leq$ $a$ $\leq$ 1. 
Evidently, a 
new  degree of freedom associated with the continuous variable $a$ 
is now present. The energy shift $E_{on}$ $\rightarrow$ 
$E_n$ is by definition adiabatical and reversible as far as 
the populations of $E_{on}$ and of $E_n$ are equal. When  
a large fraction of the initial 
energy has been transferred to $E$ levels not coinciding 
with any $E_n$, the process is not reversible anymore. 

This irreversibility can be quantified by the increase 
of the total entropy $S$. 
As a rough estimate, 
an added continuous degree of freedom plays a role 
if the total entropy increases by 
about one Boltzmann constant unit $k_b$, since in presence of $n$ 
independent degrees  
of freedom, each associated with an effective phase space 
$\Phi_i$ we have 
$S$ $=$ 
$k_B log(\prod \Phi_i/h)$ $=$ 
$k_B \sum log(\Phi_i/h)$ $\sim$ $n k_B <log(\Phi/h)>$ where 
$h$ is the Planck's constant and 
$<log(\Phi/h)>$ 
is an average phase space (since now, we will use 
$h/2\pi$ $=$ $k_B$ $=$ 1, and not explicitate these constants 
anymore). ``Effective'' phase space means that this 
phase space is at least potentially able to  
get energy within the time scale for the process interesting us. 

So,  
$\Delta S$ $\sim$ 1 may be considered a signature of the correlator 
decay, and a condition 
for a T-odd process to be present. This neither guarantees 
that the effect is a leading 
twist one, nor that its magnitude makes it observable. So 
this condition must be made more precise. 

In addition, the entropy must be defined for the relevant cases 
interesting us. 
The entropy of a slowly evolving system is defined as the entropy 
of an equilibrium system instantaneously assuming the same 
configuration. 
For a fastly evolving and open system a general definition of entropy is 
often specific and largely heuristic 
(see e.g. \cite{Kinchin57,Henson82,Zaslavsky98}) since the general 
Boltzmann's  
definitions $S$ $=$ $log(\Phi)$ or equivalently 
$S$ $=$ $-\sum p_ilog(p_i)$ face the problem of 
what must be meant by phase space $\Phi$ or 
subset probability $p_i$. 

\subsection{Time scale and non-adiabaticity} 

In practical applications, the time scale of the process is 
important: if the process is slow enough, the evolution is 
quasi-adiabatic, and relevant entropy increase may 
be meaningless since it takes place over a time range that is 
too large to influence our experiment:  
If $\tau'$ is the time 
required for reaching $\Delta S$ $\sim$ 1, and $\tau$ is the relevant 
time scale for 
the considered process, $\Delta S(\tau)$ $\approx$ $\tau/\tau'$. If 
$\tau/\tau'$ $<<$ 1, $\Delta S(\tau)$ $<<$ 1. 

So, the condition for the irreversibility of a process to play a 
role is $\Delta S(\tau)$ $\gtrsim$ 1, or equivalently 
\begin{equation}
{{dS} \over {dt}}\ \gtrsim\ {1 \over \tau}.
\label{eq:dsdt1}
\end{equation}

\subsection{Hard inelastic collisions} 

Writing eq.\ref{eq:selfcorrelation1}  
in the interaction form 
\begin{equation}
C(t)\ =\ 
\int dx {\psi_o}^*(x,t)exp\Bigg(-i\int V dt\Bigg)\psi_o(x,0), 
\label{eq:selfcorrelation2}
\end{equation}
the similarity 
between the time-dependent 
self-correlation function $C(t)$ 
and the correlation 
function entering the definition of a partonic distribution 
function\cite{Collins02} is evident, with the gauge link operator   
substituting the interaction operator. 

In a lightcone formalism 
$t$ is substituted 
by a light-cone variable and the hamiltonian by the corresponding 
conjugate light-cone momentum. 
Since here we focus on the $A(+)$ area of the factorized diagram, 
$z_-$ and $k_+$, or equivalently $\xi$ and $x$, 
play the role of time and energy. 

Because of the fourier transform 
$exp(-ix\xi)$, $\xi$-ranges  
$\sim$ $1/x$ are relevant in the problem. So, in our case 
the parameter $\tau$ of eq.\ref{eq:dsdt1} 
becomes $1/x$.   

The condition $dS/dt$ $\gtrsim$ $1/\tau$ becomes 
\begin{equation}
{{dS} \over {d\xi}}\ =\ O(x),\ for\  P_+\ \rightarrow\ \infty. 
\label{eq:dsdxi1}
\end{equation}
or more simply, for the relevant valence region 
\begin{equation}
{{dS} \over {d\xi}}\ \gtrsim\ 1,\ for\  P_+\ \rightarrow\ \infty. 
\label{eq:dsdxi1b}
\end{equation}
If $dS/d\xi$ is e.g. 0.05 this will not create problems 
(but perhaps lead to too small asymmetries to be detected). 
The point is that we 
need a visibly $finite$ limit for $dS/d\xi$ at infinite $P_+$. 

When expressed in terms of non-scaled variables, 
eq.\ref{eq:dsdxi1b} 
means 
$dS/dt$ $\sim$ $1/P_+$, so it is a singular condition 
for $P_+$ $\rightarrow$ $\infty$. 
If however this singularity 
is not present, we have entropy increase 
(that in a highly inclusive interaction is obvious) but 
at a rate that becomes adiabatic for 
$P_+$ $\rightarrow$ $\infty$. In other words, we have a 
higher twist process,  
i.e. a process that contains an intrinsic finite time decay scale 
$\tau$ and is only  
visible up to a hardness scale 
$P_+$ $\lesssim$ $1/\tau$. 

A comparison of eq.\ref{eq:gamma1}  with eq.\ref{eq:dsdxi1}  
suggests that we have to expect 
\begin{equation}
{{dS} \over {d\xi}}\ \sim\ \gamma. \label{eq:dsdxi2}
\end{equation}
The parameter $\gamma$ and the entropy $S$ have been discussed, up 
to now, as unrelated variables. 
In specific contexts we will show later that they are normally 
related by equations that imply the previous eq.\ref{eq:dsdxi2}. 

\subsection{Action of a probabilistic perturbation on a pure state}

In the following we will often consider (partially) 
probabilistic processes 
destroying an initially pure quantum state. A relevant point is 
that despite the action of this perturbation produces a chaotic 
set of states, the evolution of the $initial$ state is not cahotic. 
For an extensive and rigorous consideration of this problem in 
nuclear physics see 
ref\cite{Feshbach}. At the simple level interesting us we 
may observe that e.g. a plane wave $exp(-ipz)$ describing a 
particle crossing homogeneous nuclear matter is converted into 
a damped wave $exp(-ipz-z/L)$. Despite the damping factor 
$exp(-z/L)$ is normally estimated by means of probabilistic 
considerations, the wave $exp(-ipz-z/L)$ is a $coherent$ 
overlap of plane waves $\int dk f(k) exp(-ikz)$ with $f(k)$ 
relevant over a range $\Delta k$ $\sim$ $1/L$. 


\section{Breaking of a long range correlation} 

A point that is peculiar to this work is the hypothesis that 
the breaking of a $long$ range correlation is able to cause 
relevant $short$ range 
modifications of the quark propagators. In other words, that 
$\gamma$ in eq.\ref{eq:gamma1} represents the inverse lifetime of a long 
range hadronic state. 

To avoid misunderstandings, we stress that short range processes 
are here considered 
decisive, but their main role is to destroy long range correlations. 
The author cannot exclude that they also influence short range 
correlations. But it is more difficult to imagine processes 
where a correlation extending to a spacetime scale $L$ is destroyed 
by a mechanism with a similar spacetime scale $L'$ 
$\approx$ $L$. Also at intuition level, such 
a mechanism is able to change the underlying structure, but not 
to introduce stochasticity within a 
range $L''$ $\lesssim$ $L'$ $\approx$ $L$. 

Apart for the above general considerations, 
this choice is born from experimental data, and  
from an analysis of available theoretical models. 

\subsection{Propagation of partonic and hadronic states through 
hadronic matter} 

As already discussed in the Introduction section, 
for the Drell-Yan case there is full evidence 
that the average free path of parton distributions in nuclear matter 
is much longer 
than the corresponding path for hadronic states. 
This leads to two slightly paradoxical 
conclusions:  

(i) Short range correlations are softer than 
long range ones, when decay properties in hadronic matter 
are considered. 

(ii) $\Psi$ is destroyed while $\vert \Psi \vert^2$ 
survives during crossing hadronic matter. 

The former statement just translates the quoted experimental 
fact that both partonic distributions and their normalization 
seem to be reasonably untouched by nuclear matter. 
The second statement is necessary not to give up with 
the idea that parton distribution functions reproduce 
intrinsic, interaction independent, 
steady properties of the hadron structure. 

Putting the above pieces together into a physical picture, 
we may imagine the full process through 3 stages (the 
real inclusive process, not the one of the unitarity diagram): 

A) Long before the interaction, the wavefunction describing a single 
hadron 
has stationary features, and axial isotropy around the spin axis. 
We may speak of ``hadronic phase''. 

B) hadron-hadron interactions 
start filtering the different components of the initial 
state. Stationarity of these components is lost, and they 
are projected onto a set of short-range states that may be considered 
stationary in the crossed hadronic medium. 

C) The initial hadron coherent state is reduced to a cloud of 
individual partons easily diffusing through the hadronic/nuclear 
matter. Several probabilistic features 
of the initial state are still present, 
but fast time evolution is over. We may speak 
of ``partonic phase''.

Seen this way, the evolution is similar to the sublimation 
of a piece of solid matter passing through a high-temperature 
region. 
In the following we will not appeal to this picture anymore 
since, as above observed, we work on the imaginary part of 
the forward amplitude and 
not directly on the inclusive process. However 
this picture is 
useful to understand the role that is here attributed to 
long range correlations. 

\subsection{Breaking of long range correlations in spectator models}

The breaking of a long range correlation is a 
common point to several of the above quoted 
models for T-odd 
functions. 

In particular we may consider spectator models 
\cite{GambergGoldsteinOganessyan03,BoerBrodskyHwang03,
BacchettaSchaeferYang04,LuMa04} as an example. 

In these models, 
long range correlations are introduced and 
tuned\cite{JakobMuldersRodriguez97} 
by (i) form factors with soft cutoffs, 
(ii) unicity, compactness and stability of the spectator 
state, (iii) quark/spectator masses.  
T-odd functions may then be obtained via inclusive 
production/exchange of extra particles, that break long 
range correlations. 

In absence of extra production mechanisms, a spectator 
or a quark may correlate two 
vertexes over a distance 
$\Delta \xi$ 
$\sim$ $1/\Lambda$, 
if 
a form factor with soft cutoff $\Lambda$ is present in one 
of the vertexes or in both.  
Typically 
this means $\Delta \xi$ $\sim$ 1, since this is what is needed 
to produce a valence-like $x-$distribution with 
size $\Delta x$ $\sim$ $1 / \Delta \xi$ $\sim$ 1. 

When production/exchange of stable extra particles 
with pointlike vertex (e.g. gluons) 
is added to the basic spectator model, this 
introduces hard components 
in the quark or spectator propagator. These propagators must 
begin or end in 
a soft vertex, but the hard components cannot be reabsorbed 
by this vertex. So the projection of the initial 
hadronic state 
$\vert P \rangle$ onto the final state $\langle P \vert$ 
is suppressed by this evolution. 

This suppression is equivalent 
to the previously discussed 
breaking of long range correlations. 
A similar pattern is also present in \cite{Yuan03}, where long range 
paramenters are introduced by a Bag model. 
If the extra particles are unstable or associated with 
further vertex form factors, the 
basic features of the hadron model may play 
a minor role and the relevant 
long range correlations may be those associated with these 
fields. But the general working principle is not 
different, if the introduced parameters (inverse lifetime, 
cutoffs) don't overcome a few hundreds MeV. 

\subsection{Hard scale loops and the Fock-Krilov theorem}

In general, the reason why a loop going up to a hard 
scale $Q$ destroys 
long range correlations is a peculiar aplication of the quoted 
Fock-Krilov theorem. 
A long range correlation is associated to a soft cutoff 
$\Lambda$ on some momentum component $k$ (e.g. $k_T$) of the 
set of particles crossing the cut, in the unperturbed diagram. 
When a factorization breaking exchange introduces a loop, momenta 
through this loop go up to $Q$ $>>$ $\Lambda$. So 
the loop decomposes any wavepacket into a set of components 
with momentum $p$, each with $p$ $\lesssim$ $Q$. 
These components propagate with a different phase, so the initial 
wavepacket shape is soon lost. 
The decomposition is coherent, so sooner or later the wavepacket 
will acquire again a shape that is similar to the original 
one. Introducing the quantum state size $h$ we see that this time 
is finite since the decomposition is not really continuous. If 
however $Q$ $>>$ $\Lambda$, the number of $p-$states available 
for each given $k-$state is very large, and it is proper to 
assume that the loop variable is continuous. This however means 
that the time needed 
to reform the original wavepacket is infinite. 

At a qualitative level the increase of phase space (associated with 
an increase of entropy) associated with this loop 
is of magnitude $(Q/\Lambda)^n$, where 
$n$ is a number of order unity. 

This argument however breaks down when the ratio $\Lambda/Q$ is not 
$<<$ 1. 
In this case the pre-loop set of states composing the wavepacket 
has the same extension of the in-loop one. 
This suggests that adding further loops would not 
change enormously the effect obtained via a single hard loop.  
If in addition 
we select peculiar initial states characterized by a 
hard scale ($x$ very close to 1, or $q_T$ $\sim$ $Q$), 
decorrelation effects are suppressed. The case $x$ $\approx$ 1 is 
obvious, since we are selecting final states where decorrelation 
has not taken place at all. The large$-q_T$ case deserves more 
discussion. 

\subsection{Large transverse momenta}

In Drell-Yan at large enough $q_T$ 
the electromagnetic probe is not testing 
initial state features, but features of the hard interactions 
themselves. As a consequence it is possible to calculate 
T-odd effects\cite{JQVY06} in a fully perturbative scheme. 
As a further consequence, any correlation-destroying process 
is acting on a state that is short-ranged $\sim$ $1/q_T$ 
also in absence of factorization breaking interactions. 
So the real increase of phase space 
associated with an added factorization breaking loop 
is $(Q/q_T)^n$, not $(Q/\Lambda)^n$ as above stated. 
On the ground of the previous discussion, for  
$q_T$ $\rightarrow$ $Q$ we expect a suppression of T-odd 
effects. 
This means that in eqs. \ref{eq:damped_pole} and \ref{eq:gamma1}
$\gamma$ will not be $O(1)$ or $O(x)$ 
at large $q_T$ $\sim$ $Q$. 

In ref.\cite{JQVY06} the transition from the soft$-q_T$ to 
the hard$-q_T$ regime is considered, by including 
real gluon emission already in diagrams where factorization 
breaking loops are absent. 
For small $q_T$ the extra gluon is soft, and despite it  
enters the K-factor, it does not 
destroy the coherent features of the initial state. In this case 
the addition of a factorization breaking loop has the consequences 
described in the previous two subsections. 
For large $q_T$ $\sim$ $Q$ the 
soft features of the initial state are destroyed by the 
hard radiated gluon. In other words, also in diagrams with 
no factorization breaking interactions, initial state soft 
cutoffs play no role. 

So, in the case of the models considered in the previous 
subsections V.B and V.C, and for small $q_T$ we may speak of 
``soft to hard'' transition. In the case of large $q_T$ 
we may speak of ``hard to hard'' transition, with no 
decorrelation. This reflects in a $q_T-$suppression of the 
effect at large $q_T$, as predicted in\cite{JQVY06}. 


\subsection{Soft cutoffs}

Let a variable $Y$ 
be connected with the quark hole motion. It may represent 
the quark transverse momentum, or be related with gluons 
radiated by this quark, and so on. 

Let $Y$ have a maximum ``soft''  
cutoff $Y$ $<$ $Y_M$ in its initial state. 
Let the quark hole 
undergo events associated with 
initial/final state interactions\footnote
{E.g., in a spectator model a gluon production from a quark 
propagator is an initial state interaction, a gluon 
production from a diquark line is a final state interaction.}, 
that we simply 
name ``events''.

We assume that most events lead to 
$Y$ $<$ $Y_M$ (``soft'' events), 
but for a small and finite fraction 
of events 
$Y$ $>>$ $Y_M$ (``hard'' events). 

To avoid ambiguities, we stress that 
in the following we will speak of sof and hard events 
referring, in both cases, to 
initial/final state interactions 
that accompany the electromagnetic Drell-Yan hard event, not 
to the electromagnetic event itself. 

The meaning of the soft cutoff is a sharp way to represent 
a reasonable upper cutoff for all fields and degrees of freedom 
that may play 
a role in the hadron state 
when $P_+$ $\rightarrow$ $\infty$. The assumption is that if 
$Y$ $<$ $Y_M$ the hadron is in its ground state, if 
$Y$ $>$ $Y_M$ it is in some excited state, if $Y$ $>>$ $Y_M$ 
it is fragmenting. 
Since the relevant correlation 
function is defined as $\langle P \vert ... \vert P \rangle$, i.e. 
as a correlation operator sandwiched between 
equal hadron states, 
if some internal degree of freedom largely overcomes 
its soft cutoff the 
projection on the final 
$\langle P \vert$ state is very small. 
We may imagine some relevant cases: (i) 
$Y_M$ coincides with the vacuum expectation value for $Y$, (ii) 
$Y_M$ is assigned by the uncertainty principle (e.g. 
for $k_T$ we have $max(k_T)$ $\sim$ $1/R_{hadron}$),  
(iii) $Y_M$ is an infrared cutoff. 

A relevant point is that 
\begin{equation}
S\ \approx\ 0, \hspace{0.5truecm} for\ Y\ <\ Y_M. \label{eq:s_ground} 
\end{equation}
Indeed, as just observed, as far as 
the soft variables are inside their 
soft cutoffs the considered state is 
a quantum fluctuation of the hadron ground state. 

\subsection{Single and multiple step processes: hard and stretching events}

A series of events may determine a transition of the bound hadronic state 
to the continuum. We expect that the entropy is then increased 
by some units (or more), and correspondingly that 
some more degrees of freedom have appeared. 
This may take place within a single or a multiple scattering scheme. 

If this process is 
dominated by a single hard event, we need the  
final phase space of this single event to contain 
more degrees of freedom than the initial one. This can be obtained 
in several ways: 
(i) by infinitely 
expanding the phase space of an existing degree of freedom;  
(ii) by producing some new particle (that is a subclass of the previous 
case, since the associated phase space expands from zero to a finite 
value); 
(iii) for the case of an initially discrete set of values, 
by filling each range $Y_n$ $-$ $Y_{n-1}$ 
with an infinite set of new accessible values (that again is 
a subclass of (i)). So, essentially a single step process 
is dominated by an infinite expansion of the phase space.\footnote{ 
If the phase space is 
evalued in classical sense, i.e. without counting the finite volume 
$h$ of each state. This is justified since we reach quasi-classical 
situations, where $h$ may be neglected. 
}

In a multistep process associated with a large number of uncorrelated 
events, it may be sufficient to have 
a regular increase of the phase space associated with a given  
variable at each step. This wll lead to an 
exponential increase of the phase space with the event number, 
and the effect on $S$ is still an increase of some units.  
In this case we speak of ``stretching events'' instead 
of ``hard events''. 

Single and multiple step processes are separately 
examined in the following. 
Hybrid features do normally appear,  
e.g. when a hard scattering 
leads to a multiparticle production, or 
when in a multiparticle production we may  
radiate a few hard particles and several 
soft ones as well. 
So the above two sets of processes should 
be considered just as opposite limits for a continuum of 
possibilities. In these limits approximations may be applied 
making statistical analysis easier. 

In the format given by ref.\cite{Collins02} and widely 
adopted 
as a formal framework for adding the effects of 
initial state interactions to 
partonic distributions, the gauge link factor 
incorporates interactions in 
a continuous and abstract form. So there is no difference in principle 
between the two classes of events. 
The problem 
rises when the gauge factor is expanded in a perturbative  
scheme, and one needs to decide whether  
low order diagrams, or a full 
resummation scheme, have more relevance. 

For a specific model it may be easier to discuss 
low order perturbative schemes. Sometimes however 
resummation is hidden, in form factors, in finite widths 
for propagators, in the use of no-event probabilities. 
E.g. the factor 
$exp(-\gamma \xi)$ used in several places of this work is 
a quasi-classical 
all-order resummation of uncorrelated similar events. 

\subsection{Single scattering/radiation scheme: 
real and effective degrees of freedom} 

In the limit $P_+$ $\rightarrow$ $\infty$, in presence of a 
hard exchange of energy or transverse momentum between the 
factorization separated areas of the diagram $A(+)$ and $A(-)$, 
large amount of energy can be reversed on a set of soft variables 
$\{Y_i\}$ 
belonging to $A(+)$. As above anticipated, we may 
select three typical situations, that actually are specializations 
of the first one. 

We use a single variable $Y$ to represent 
a combination of all the involved ones. E.g., it may be their total 
energy. $Y$ $<$ $Y_M$ in soft conditions. 

If hard events may lead 
to a value for $Y$ distributed in a reasonably uniform way 
within a range $\Delta Y$ with $\Delta Y/Y_M$ 
$\rightarrow$ $\infty$ for $P_+$ $\rightarrow$ $\infty$,  
and if hard and soft events coexist, 
we need $two$ continuous variables $a$ and $y$ 
to describe the 
final state. 

Let $Inf(z)$ be a monotonously increasing 
function of $z$ that is infinite for infinite 
$z$ and positive for $z$ $>$ 1. 
$Log(z)$, $z$, or $exp(z)$ are all good functions of this kind. 

We define 
\begin{equation}
Y\ \equiv\ a Y_M\ +\ y Y_M Inf(P_+), \hspace{0.5truecm}
0\ \leq\ a,y\ \leq\ 1. 
\label{eq:y1}
\end{equation}
In the limit $P_+$ $\rightarrow$ $\infty$ we have 

\noindent
-) soft events: $a$ has a precise value between 0 and 1, $y$ $\equiv$ 0.

\noindent
-) hard events: 0 $\leq$ $y$ $\leq$ 1, 
$a$ may be arbitrarily assigned within 0 and 1. 

\noindent
For hard events $a$ would be infinite if we had defined 
$Y$ $=$ $a Y_M$ without introducing $y$. 
So, the joint presence of soft and hard events 
(i.e. of $Y$ values ranging over two infinitely different scales 
of magnitude) 
implies the 
effective opening of a new degree of freedom. 

When the soft cutoff coincides with the vacuum expectation 
for $Y$, we may roughly 
speak of a ``new turned on'' degree of freedom. 
In this case we may simply define: 

\begin{equation}
Y\ \equiv\ Y_{vacuum}\ +\ y Y_o, \hspace{0.5truecm} y\ \sim\ 1. 
\label{eq:y2}
\end{equation}
where the constant $Y_o$ keeps $y$ dimensionless. 
This is a special case of the previous definition, with 
$Y_M$ $\equiv$ $1/Inf(P_+)$ and 
the soft phase space reduced to unity. 

Another case is the one of a variable 
with a discrete set of eigenvalues $Y_n$ below the cutoff, and 
a continuous set over the cutoff. If the number of discrete 
eigenvalues is finite this case is equivalent 
to the previous one. If it is infinite we may map them 
univoquely into a set of discrete $Y_n'$ eigenvalues 
spanning all the final state space. Then we may define 

\begin{equation}
Y\ \equiv\ Y_n'\ +\ y (Y_n' - Y_{n+1}') , \ 0\ <\ y\ <\ 1. 
\label{eq:y3}
\end{equation}

Other cases that one may 
imagine may be reduced to the discussed ones. 
In all cases one sees that a new continuous variable $y$ 
has been introduced. 

One important remark: In the previous cases two dimension 
scales are selected for the same set of events.  
Separating the associated phase space into two degrees of freedom 
$a$ and $y$ shows that, despite appearances, the final phase space is 
not infinitely large. It only has a different dimension. 
Since the two final 
degrees of freedom are roughly independent, we have 
$S$ $\sim$ $log(\Phi(a)\Phi(y))$ $\sim$ $S(a)$ $+$ $S(y)$, 
where $\Phi(a)$ and $\Phi(y)$ are the phase spaces 
associated with each variable.     
The former quantity is zero because of eq.\ref{eq:s_ground}, and the 
latter is $\sim$ 1.\footnote{It may be much smaller ot much larger, but 
neither zero nor infinite; its precise value depends on our 
ability in distinguishing different $y$ values within a distance 
$\delta y$; then $S(y)$ $=$ $-log(\delta y)$.}

\subsection{$dS/d\xi$ in a single scattering/radiation  
scheme} 

Let us first consider a single scattering center 
in the unit volume. 
For the probability $W$ of transitions of both soft and hard kinds, 
we have 

\begin{equation}
W\ =\ W_{soft}\ +\ W_{hard}\ 
\approx\ \vert M\vert^2 \Big(\Phi(a)\ +\ n\Phi(a,y)\Big)\ 
\approx\ \vert M\vert^2 \Big(\Phi(a)\ +\ n\Phi(a)\eta_y\Big), 
\label{eq:W1}
\end{equation}

\begin{equation}
\eta_y\ \equiv\ \Phi(a,y)/\Phi(a).  
\label{eq:etay1}
\end{equation}

In the previous relations $n$ is the relative fraction of hard to 
soft scattering centers. ``Scattering centers'' means 
partons in $A(-)$ potentially able to cause one of the required 
processes. 
$\Phi(a)$ and $\Phi(a,y)$ are 
the phase space 
associated with soft transitions and hard transitions 
respectively. 
$\vert M\vert^2$ is an average soft transition probability to a single 
final state. The corresponding $\vert M_{hard}\vert^2$ 
factor for hard transitions is supposed 
to differ mainly for selection rules (included in $n$) and 
for conservation rules (included in the phase space). So 
what remains in 
$\vert M \vert ^2$ is supposed to be similar for soft and hard 
transitions. 

We need to be inclusive with respect to the initial values 
of the soft variable $a$, so the overall transition probability must 
be divided by the soft phase space $\Phi(a)$. So we have 

\begin{equation}
W\ =\ W_{soft}\ +\ W_{hard}\ 
\approx\ \vert M\vert^2 \Big(1\ +\ n\eta_y\Big)\ 
\equiv\ \vert M\vert^2 \Phi_{tot}\ 
\equiv\ \vert M\vert^2 exp(S). 
\label{eq:W2}
\end{equation}

The previous equation may be considered as the definition 
of entropy that is relevant here: the logarithm of the number 
of all the final states that directly determine the rate of 
hard interactions of the quark hole, starting from a soft 
state. 

This excludes all those processes that are present but have scarce 
relation with the quark hole line. With the above definition of  
uesful phase space, we have $S$ $=$ 0 for soft final states, 
as previously required. 
Of course, deciding what must be counted in the relevant phase 
space is a matter of taste, but this problem is present 
in the calculation of any inclusive reaction rate in hadron physics. 

Assuming that relevant values for $S$ and $n\eta_y$ 
are $\lesssim$ 1, 

\begin{equation}
e^S\ \approx\ 1\ +\ S\ \approx\ 1\ +\ n\eta_y, 
\label{eq:S1}
\end{equation}

\begin{equation}
W_{hard}\ 
\approx\ W_{soft} S, \hspace{0.5truecm} S\ \approx\ n\eta_y
\label{eq:S2}
\end{equation}

If we consider events taking place with continuity along a $\xi$ path, 
the previous equations need to be modified: 

\begin{equation}
{{dW} \over {d\xi}} 
\ =\ 
\Bigg({{dW} \over {d\xi}}\Bigg)_{soft}\ +\ 
\Bigg({{dW} \over {d\xi}}\Bigg)_{hard}\ \approx\ 
\Bigg({{dW} \over {d\xi}}\Bigg)_{soft}(1\ +\ S)
\label{eq:W3}
\end{equation}

If with $dW_{soft}/d\xi$ we mean the full set of processes 
that leave a soft state untouched after a unitary path $\Delta \xi$ 
$=$ 1, we may approximate it with 1 (since scattering and no-scattering 
events are not distinguishable). Then $n$ (the $relative$ number of 
hard to soft scatterers) must be substituted by 
$dn/d\xi$, where $dn/d\xi$ is the total $absolute$ number 
of hard scatterers per unitary $\Delta \xi$, 
and consequently $S$ $\rightarrow$ $dS/d\xi$: 

\begin{equation}
\Bigg({{dW} \over {d\xi}}\Bigg)_{hard}\ \approx\ 
{{dS} \over {d\xi}}
\label{eq:dwsxi1}
\end{equation}

\begin{equation}
{{dS} \over {d\xi}}\ \approx\ 
{{dn} \over {d\xi}}\eta_y 
\label{eq:dsdxi3}
\end{equation}

Now, the probability for the set of soft states suffers unitarity loss: 

\begin{equation}
\vert \psi(x, Y < Y_M, \xi)\vert^2 
\ \sim\ \vert \psi(x, Y < Y_M, 0)\vert^2 e^{-2\gamma \xi},  
\label{eq:gamma3}
\end{equation} 
since for the overall flux 
we have $dN/N$ $=$ $(dW_{hard}/d\xi)d\xi$, and $dW_{soft}$ 
does not contribute to the decay.  
So we get the relation between $dS/d\xi$ and $\gamma$ 
whose approximate form (eq.\ref{eq:dsdxi2})  
was guessed in section IV: 
\begin{equation}
\gamma\ =\  {1 \over 2} {{dS} \over {d\xi}}. 
\label{eq:gamma4}
\end{equation}
and 
\begin{equation}
{{dS} \over {d\xi}}\ \sim\ x
\label{eq:dsdxi4}
\end{equation}
that selects leading twist T-odd effects. 

Eq.\ref{eq:dsdxi3}  suggests that 
a finite $dS/d\xi$ 
may be reached more than one way: 

1) Both $dn/d\xi$ and $\eta_y$ finite for $P_+$ $\rightarrow$ $\infty$. 

2) $dn/d\xi$ $\rightarrow$ $\infty$ and 
and $\eta_y$ $\rightarrow$ 0 for $P_+$ $\rightarrow$ $\infty$. 

When $\xi$ is converted to the non-scaled variable $z_-$, the less 
singular case may be the latter, since a unit $\xi$ corresponds to 
an infinite $z_-$ range, and a finite $\eta_y$ implies infinite 
energy being acquired by $Y$ in the limit 
$P_+$ $\rightarrow$ $\infty$. However, in this limit it is quite 
natural to have to do with singularities. 

In any case, both 
the above schemes are single scattering schemes, meaning 
that a quark hole flux moves in a medium with several potential 
scatterers, but one event is sufficient for the destruction of the 
ground hadron state. 

\subsection{Stretching events and 
multiple scattering/radiation schemes for $dS/d\xi$} 

A further scheme is given by multiple 
uncorrelated scattering/radiation events.  

A known example of this kind that could be generalized 
to our problem is a classical particle 
undergoing multiple scattering against randomly distributed 
spheres\cite{Krilov}. Then  
$Y$ could be the angular deviation of the quark from 
the hadron direction. 
Next, this class of processes 
includes relevant cases like 
radiation of gluons/mesons by a quark line. Then 
$\{Y_i\}$ may be 
the set of kinematic variables of the radiated particles. 

In a multiple scattering/radiation case, 
hardness is reached 
as a cumulative effect of several events that may be individually 
soft. This obliges us to modify the 
definition of $dn/d\xi$. 

Let us redefine $Y$ in such a way that it has the dimension of an 
action or, if a unitary cubic box has been assumed as normalization 
volume, 
of a momentum. We scale 
$y$ $\equiv$ $Y/Y_M$, with $y$ 
that is soft ($<$ 1) 
at the very beginning, but may reach any value during the evolution 
driven by initial/final state interactions. 
If the values of $y$ at a certain time are 
randomly distributed within a mean quadratic 
fluctuation, $y \pm \Delta y$, the phase space that we may 
associate to $y$ is a simple power $m$ of $\Delta y$. 
Consequently, 

\begin{equation}
S\ \equiv\ m\ log(\Delta y),\hspace{0.5truecm} m\ \sim\ 1. 
\label{eq:multiple1}
\end{equation}

Instead of defining $dn/d\xi$ as a 
density of scattering centers for hard events, 
we define now $dn/d\xi$ as a density of 
scattering centers for ``stretching events''. A  
stretching event is defined by the condition: 

\begin{equation}
{{\Delta y'} \over {\Delta y}}\ 
\equiv\ 1\ +\ \alpha_y, \ \alpha_y\  
>\ 0, \ for\ P_+\ \rightarrow\ \infty. 
\label{eq:multiple2}
\end{equation}
where $\Delta y$ is the mean quadratic spread for $y$ 
values before the event, and $\Delta y'$ the same after 
it. The phase space factor $\alpha_y$ includes in itself the 
total probability 
of scattering in presence of a single scattering center. 

In several problems, 
$\alpha_y$ $>$ 0 is valid at finite $P_+$, 
but it is lost in the $P_+$ $\rightarrow$ $\infty$ 
limit. As an example 
we take  $y$ as the scattering angle in 
Rutherford scattering. 
Assuming an upper cutoff on the impact parameters $z_T$ $<$ 
$z_{max}$, we have 
$\alpha_y$ $>$ 0 for finite $P_+$, but  $\alpha_y$ 
$\rightarrow$ 0 in the $P_+$ $\rightarrow$ $\infty$ limit.\footnote{
If no upper limit is assumed for $z_T$ the average scattering angle 
is zero at any energy.}

Assuming that a finite set of events respect the stretching condition, 
may define $S$ as  

\begin{equation}
S\ =\ log(\Delta Y / \Delta Y_o), 
\label{eq:multiple3}
\end{equation}
where $\Delta Y_o$ is a ``soft'' initial value. With this definition 
$S$ is proportional 
to the number of scattering events. Using all quantities referring to 
one $\xi$ units, and neglecting the precise value of $m$ in 
eq.\ref{eq:multiple1}, 

\begin{equation}
{{dS} \over {d\xi}}\ \approx\ log(\Delta y)_{\Delta \xi = 1}\ 
\ =\ log[\Delta y_o\ (1 + \alpha_y)^{(dn/d\xi)}] \ 
=\ {{dn} \over {d\xi}} log(1+\alpha_y). 
\label{eq:dsdxi5}
\end{equation}

Assuming a small $\alpha_y$ we have 

\begin{equation}
{{dS} \over {d\xi}}\ \approx\ 
{{dn} \over {d\xi}} \alpha_y.
\label{eq:dsdxi6}
\end{equation}

In the hadron ground state we have $S$ $\approx$ 0, as a result of 
having neglected $log(\Delta Y_o)$ in eq.\ref{eq:dsdxi5}. 
To estimate the relation between $\gamma$ and $dS/d\xi$ we may 
use the known trick of considering a set of $j$ independent 
emissions in a range $\xi$ 
as Poisson-distributed with respect to $j$, 
and then $exp(-2\gamma\xi)$ is identified as the probability of 
no-emission. In the Poisson distribution this is 
$exp(-\omega_1)$, where $\omega_1$ is the probability 
of a single emission. Since here with ``emission'' we mean a 
process with increase of phase space, we may identify 
$(1 + \alpha_y) dn/d\xi$ with the probability of an emission 
of whatever kind in a unit $\xi$ range, 
and $\alpha_y dn/d\xi$ with the probability for 
a ``useful'' emission only, since the other processes don't 
change the initial situation. So 

\begin{equation}
exp(-2\gamma \xi)\ \sim\ 
exp\Bigg(-\alpha_y {{dn}\over {d\xi}} \xi\Bigg)\ =\ 
exp\Bigg(-{{dS}\over {d\xi}} \xi\Bigg)
\label{eq:dampedmultiple1}
\end{equation}

So, again we have $\gamma$ $\sim$ $dS/d\xi$ and consequently 
the requirement $dS/d\xi$ $\sim$ $x$. 
In addition $\alpha_y$ plays a similar role as  
$\eta_y$ in the single scattering scheme. 

Also in this case we may have two possibilities: 

3) Both $dn/d\xi$ and $\alpha_y$ finite for $P_+$ $\rightarrow$ $\infty$. 

4) $dn/d\xi$ $\rightarrow$ $\infty$ and 
and $\alpha_y$ $\rightarrow$ 0 for $P_+$ $\rightarrow$ $\infty$.

\section{T-odd functions in 
energy conserving schemes} 

In the title of this section ``energy conserving'' means 
that initial/final state interactions don't exchange energy 
between factorization separated areas of the diagram. This is 
a natural consequence of relating T-odd effects 
with entropy nonconservation in $A(+)$. 
Entropy may increase also in 
energy-conserving systems. Since 
$O(P_+)$ energy exchanges are statistically 
suppressed in hadron-hadron collisions at high energies,  
the relevance of this possibility is evident. 

As observed in section III, 
T-odd functions describe loose of flux due to 
events where $x$ is not conserved in the propagation 
of the quark hole, with finite $x$ $-$ $x'$ 
for $P_+$ $\rightarrow$ $\infty$. 
Since the colliding hadron is a stationary system if undisturbed, 
$x$ nonconservation 
takes place because of hard interactions 
between $A(+)$ and $A(-)$. 
We must however distinguish 
between two very different 
classes of hard events: 

1) Energy nonconserving events:  $O(P_+)$ energy 
exchanges take place between $A(+)$ and $A(-)$,   
and the lost energy is not returned. 

2) Events where the hard interaction between 
$A(+)$ and $A(-)$ is mainly transverse, it does 
not transfer relevant amounts of energy between the two, 
and only act    
as a ``trigger'' for relevant energy exchanges between 
degrees of freedom 
all belonging to $A(+)$. 
In particular, energy is transferred 
from $x$ 
to variables that consequently 
overcome the soft cutoff. 

In this section, we focus on events of the latter kind. 
These necessarily increase the entropy of $A(+)$ and 
correspond to what, in classical mechanics, 
is energy degradation: energy is transferred from 
``mechanical'' to ``internal'' degrees of 
freedom 
(from collective, ordered, translation and rotation motion  
to thermal or chemical energy). Since the macroscopic 
dynamical evolution is associated with mechanical 
degrees of freedom only, energy degradation is equivalent 
to a energy nonconservation.  
In our problem, the mechanical degrees of freedom are 
$\xi$ (conjugated to $x$), and the quark transverse motion as 
far as $k_T$ remains within 
soft limits. 

The above requirement of energy conservation inside $A(+)$ 
puts limits to the size of the final state phase 
space and allows for some approximate estimate. 

\subsection{Soft and hard cutoffs on the phase space} 

Specializing the assumptions of the previous section V, 
we may assume that in $A(+)$ we have, for any given 
$x$, a set of $m$ 
variables $Y_1, Y_2, ..Y_m$ that in the hadron ground 
state satisfy one or more soft cutoffs 
of the form $f(Y_1, Y_2, ..)$ $<$ $f_M$. In particular this set 
contains $\vec k_T$ satisfying 
${k_x}^2$ $+$ ${k_y}^2$ $\lesssim$ ${\Lambda_T}^2$, whith $\Lambda_T$ 
of magnitude 1 GeV/c.  
The set $\{Y_i\}$ $\equiv$ $Y$ only 
includes  
those degrees of freedom that may exchange energy with the quark 
hole within a time $R_{hadron}/P_+$. 

The soft cutoff will translate, in particular, in a soft cutoff 
for the $total$ energy associated with $Y$: 

\begin{equation}
\sum E(Y_i)\ \equiv\ E(Y)\ 
\lesssim\ f_M. 
\label{eq:cutoff1}
\end{equation} 
Since the total energy $E(x)$ $+$ $E(Y)$ is conserved, 
in the hadron ground state at large $P_+$, 
also the longitudinal energy $E(x)$ 
may ordinarily fluctuate within this soft limit: 
\begin{equation}
E(x)\ -\ E(x')\ \sim\ 
(x\ -\ x')P_+\ \lesssim\ f_M, 
\label{eq:cutoff2}
\end{equation} 
so obviously soft events don't break $x-$conservation in the 
large $P_+$ limit. 

Hard events break the soft cutoff. We assume that they lead to 
a reasonably nonzero distribution for the final values 
of $Y$ up to a hard cutoff: 

\begin{equation}
E(Y)\ \lesssim\ F_M(x)\ >\ f_M. 
\label{eq:cutoff3}
\end{equation} 

\noindent 
So, for soft events we have the phase space 
\begin{equation}
x\ -\ x'\ < {f_M \over {x P_+}} 
\label{eq:cutoff4}
\end{equation} 
and 
for hard events we have the phase space 
\begin{equation}
{f_M \over P_+} <\ 
x\ -\ x'\ < {F_M \over {x P_+}} 
\label{eq:cutoff5}
\end{equation} 
that can be simply approximated by 
\begin{equation}
x\ -\ x'\ < {F_M \over {x P_+}}  
\label{eq:cutoff6}
\end{equation} 

\noindent 
We also limit our discussion to the cases where 
\begin{equation}
x\ -\ x'\ << x  
\label{eq:smallx}
\end{equation} 
This means that 
$1/x$ and $1/x'$ are interchangeable in the right hand side 
of the above equations. Clearly this is valid as far as 
$F_M$ $<<$ $xP_+$ for $P_+$ $\rightarrow$ $\infty$, 
and $x$ is not too close to zero. 

The total phase space $x'$ $\leq$ $x$ 
for one given $x$ in the case of soft/hard events is 
\begin{equation}
\Phi(E_M,x)\ =\ 
\int_0^x
{{dx'} \over {x'}} \int_{E(Y) < E_M} d^m\Phi(Y) 
\delta\Bigg(x-x'-{{E(Y)}\over {x'}}\Bigg) 
\label{eq:cutoff7}
\end{equation} 
where $E_M$ is the maximum for $E(Y)$, i.e. either $f_M$ or $F_M$. 
We may evidentiate 

\begin{equation}
d^m\Phi(Y)\ \equiv\ dE(Y)d^{m-1}\Phi(y)
\label{eq:cutoff8}
\end{equation} 
with 

\begin{equation}
y_i\ \equiv\ {{E(y_i)} \over {E_M}} 
\end{equation} 
\begin{equation}
\int d^{m-1}\Phi(y)\ \equiv\ [E(y)]^{m-1} \int \delta(1-\sum y_i)
f(x,y)\prod_i dy_i\ \equiv\ v(x) [E(y)]^{m-1}
\label{eq:cutoff9}
\end{equation} 

\noindent 
where $f(x,y)$ is supposed to be reasonably flat through all 
the $m-1$ above defined phase-space, and to depend slowly 
on $x$ as $v(x)$. 
Assuming 
\begin{equation}
{{x-x'} \over x}\ \lesssim\ 1/P_+
\label{eq:smallx2}
\end{equation} 
(that is obvious but at very small $x$ values) 
from the joint conditions $x$ $=$ $x'+E(Y)/xP_+$ and 
$E(Y)$ $<$ $E_M$ we have: 

\begin{equation}
\int_0^x
{{dx'} \over {x'}} 
\delta\Bigg(x-x'-{{E(Y)}\over {x P_+}}\Bigg) 
\approx\ 
P_+ \int_{x-\Delta}^x
dx' 
\delta\Bigg(P_+ x(x-x')-E(Y)\Bigg), \ \Delta\ =\ {E_M \over {xP_+}}. 
\label{eq:cutoff10}
\end{equation} 
and with some algebra 

\begin{equation}
\Phi(E_M,x)\ =\ 
v(x) {{E_M}^m \over m}.
\label{eq:cutoff11}
\end{equation} 

\noindent 
Taking into account that we have neglected the lower cutoff in 
the calculation of the hard phase space (eq.\ref{eq:cutoff5} has been 
approximated by eq.\ref{eq:cutoff6}), 
\begin{equation}
{{\Phi_{hard}} \over \Phi_{soft}}\ =\ 
{{\Phi(F_M,x)} \over {\Phi(f_M,x)}}\ -\ 1
\ \approx\ 
\Bigg( {F_M \over f_M} \Bigg)^m\ -\ 1.
\label{eq:cutoff11b}
\end{equation} 

In a single scattering scheme this may be identified with $\eta_y$ 
of eq.\ref{eq:etay1}. In $dS/d\xi$ also $dn/d\xi$ appears, where 
this number is the density of scattering 
centers able to trigger the energy redistribution process. 

As interesting limiting cases, we may imagine 
two situations: 

(i) $F_M/f_M$ close to 1, large $m dn/d\xi$: a multiparticle production, 
or anyway a redistribution of energy among a 
large number of degrees of freedom. 

(ii) Large $F_M/f_M$, small $m dn/d\xi$: a single particle radiation 
or anyway few involved degrees of freedom. 

\subsection{Large number of degrees of freedom} 

If an average $m$ is fixed we have 

\begin{equation}
(F_M/f_M)^m\ \equiv\ (1\ +\ \delta)^m\ \approx\ (1\ +\ m\delta), 
\ \delta\ <<\ 1.  
\label{eq:cutoff12}
\end{equation} 

\begin{equation}
{{dS(x)} \over {d\xi}}\ \sim\ 
\delta m {{dn} \over {d\xi}}.
\label{eq:dsdxi10}
\end{equation}

Else, if $m$ is Poisson-distributed, we may 
work as in the multiple scattering case (the two situations 
are now equivalent), and get the same results with the factor 
$1+\alpha_y$ of eq.\ref{eq:multiple2}  
substituted by $1+\eta_y\vert_{m=1}$: 

\begin{equation}
{{dS(x)} \over {d\xi}}\ \sim\ 
\delta {{dn} \over {d\xi}}.
\label{eq:dsdxi11}
\end{equation} 

In both cases $dS/d\xi$ is proportional to the relative  
excess of released energy $\delta$ $=$ $(F_M-f_M)/f_M$ 
times the average number of new particles produced per unitary $\xi$ 
path. The soft cutoff is linearly present in $\delta$. 

\subsection{Single particle production} 

We neglect the ``1'' factor in eq.\ref{eq:cutoff11b}. 

In this case also single particle production in 
constituent or spectator models 
may enter, but then the previous treatment of the phase space 
ratio must be corrected. Now the numbr of degrees of freedom is 
limited, and one needs to 
specify which ones play a role and where. 

In a spectator model 
the effective number of degrees of freedom that can be associated 
with the initial state 
is roughly 
$2\alpha$, where $\vert 1/Q^2 \vert^\alpha$ is the asymptotic 
electromagnetic hadron form factor as it would result when calculated 
within the considered model. In a spectator model like the one
of\cite{JakobMuldersRodriguez97} this power-law 
derives from form factors in the hadron-quark-diquark vertexes of 
the kind 
$(k^2-{m_q}^2)^a/\vert {k_T}^2+\lambda^2\vert^b$. 

If e.g. the diquark emits a gluon, the only degrees of freedom 
that may break softness are the two components of the recoiling 
diquark $\vec k_T'$. These are soft-limited by the final 
vertex form factor. As above observed, 
they are effectively treated as if they 
were $2\alpha$ instead of two. On the other side, 
the effects of the vertex form factors disappear in the 
asymmetries, where form factors appear both in the numerator 
and in the denominator. 

So in T-odd distributions we have 
\begin{equation}
\eta_y\ \propto\ \lambda^{-2\alpha}
\label{eq:cutoff20}
\end{equation} 

As a result, T-odd distributions are $\sim$ $1/\lambda^m$ with 
relevant $m$ factors. 

In asymmetries, on the other side, we only have 
\begin{equation}
\eta_y\ \propto\ \lambda^{-2}
\label{eq:cutoff21}
\end{equation} 
as one may expect on the ground of the previous counting 
rule (see e.g. \cite{BacchettaSchaeferYang04}). 

This suggests that, despite in a hidden way, in models with this scaling 
law (asymmetry $\sim$ 
$1/\lambda^2$) gluons exchanged between factorization separated 
areas of the diagram mainly carry transverse momemtum. 

As observed in section $V$, at large $q_T$, $\Lambda$ must be 
substituted by $q_T$. In addition, the power law must be reconsidered 
taking into acount the number of poles constraining the real number 
of added degres of freedom (see ref.\cite{JQVY06}). 


\section{Conclusions}

The initial state interactions affecting a 
Drell-Yan process have been reanalyzed. Assuming, 
on an experimental ground, that they 
cause a transition from a hadronic to a partonic 
phase, nonzero leading twist T-odd distribution functions 
have been shown to be closely associated to this phase 
transition. 

A parallelism has been established between the 
time-dependent self-correlation 
function describing the time evolution of a state 
in an interacting quantum system, and the 
light-cone correlator from which a distribution 
functions is extracted via Fourier transform $exp(ix\xi)$. 
The decay of the correlator is associated with entropy 
increase. 

We have derived in heuristic way the condition 
$dS/d\xi$ $\sim$ $x$. If it is respected in the infinite 
momentum limit, a leading twist T-odd structure is present in the 
correlator 
and it may be made detectable. In particular, the 
correlator is a sum of decaying functions of the kind 
$exp[-ix\xi -\gamma(x)\xi]$, and $\gamma$ $\sim$ $dS/d\xi$.  

Some general schemes for single and multiple scattering/radiation have 
been examined arriving, in each case, to relations of the kind 
$\gamma$ $\approx$ $dS/d\xi$ $\approx$ $\eta dn/d\xi$, where $\eta$ is the 
relative gain of phase 
space associated with a single interaction event, and $dn/d\xi$ 
the density of scattering centers along the quark path. 

Last, we have examined the special situation where the 
initial state interactions do not exchange energy between 
the two hadrons. For this case, that is likely to dominate 
initial state interactions 
in high energy Drell-Yan, 
the entropy increase rate $dS/d\xi$ has been estimated, in the opposite 
cases of single and multiple radiation.



\end{document}